\documentclass[preprint,dvipdfmx,groupedaddress,aps,prd,longbibliography]{revtex4-2}

\usepackage[dvipdfmx]{graphicx}
\usepackage{dcolumn}	% Align table columns on decimal point
\usepackage{bm}			% bold math
\usepackage{here}
\usepackage[dvipdfm]{}
\usepackage{graphicx}
\usepackage{mathrsfs}
\usepackage{amsmath}
\def\tc{$T_{\rm c}$}
\def\ts{$T^*$}

\def\cx{Cu$_x$Bi$_2$Se$_3$}
\def\sx{Sr$_x$Bi$_2$Se$_3$}
\def\nx{Nb$_x$Bi$_2$Se$_3$}
\def\BS{Bi$_2$Se$_3$}
\def\hct{$H_{\rm c2}$}

\begin{document}

\title{Spontaneous Lattice Distortion %and Two-component Order Parameter  %below the Superconducting Transition 
	in the Spin-Triplet Superconductor Cu$_{x}$Bi$_2$Se$_3$}

%\author{
	\author{K. Matano$^{1,2}$}   
%	\affiliation{Department of Physics, Okayama University, Okayama 700-8530, Japan}
%	\altaffiliation{Department of Physics, Okayama University of Science, Okayama, 700-0005, Japan.}%\altaffiliation{},
	\author{S. Takayanagi$^{1}$}
	\author{K. Ito$^{1}$}
	\author{S. Nita$^{1}$}
	\author{M. Yokoyama$^{1}$}
	\author{M. Mihaescu$^{1}$}
	% K. Ishii$^{2}$,
	\author{H. Nakao$^{3}$}
	%and 
	\author{Guo-qing Zheng$^{1}$}
%	\thanks{To whom correspondence should be addressed; E-mail:  zheng@psun.phys.okayama-u.ac.jp}
\email[]{zheng@psun.phys.okayama-u.ac.jp}
%}
%\email[]{Your e-mail address}%\thanks{}%\altaffiliation{}
\affiliation{
	$^1$Department of Physics, Okayama University, Okayama 700-8530, Japan\\
%	$^2$Synchrotron Radiation Research Center, National Institutes for Quantum and Radiological Science and Technology Hyogo 679-5148, Japan\\
$^2$Department of Physics, Okayama University of Science, Okayama, 700-0005, Japan.
	$^3$Photon Factory, Institute of Materials Structure Science,
	High Energy Accelerator Research Organization, Tsukuba 305-0801, Japan\\
}
\date{\today}

\begin{abstract}
The doped topological insulator Cu$_x$Bi$_2$Se$_3$ has attracted considerable attention as a new platform for studying novel properties of spin-triplet  and topological superconductivity. 
%While spontaneous in-plane symmetry breaking of superconducting properties has been examined from various perspectives,
%the relationship between the pinning of the  ${\boldsymbol d}$-vector and the crystalline lattice presents a novel issue not observed in previous spin-triplet superconductors.
In this work, we performed synchrotron x-ray diffraction measurements on Cu$_x$Bi$_2$Se$_3$ (0.24$\leq x\leq$ 0.46)  to investigate the coupling between the superconducting order parameter and  crystal lattice. % whether crystallographic distortions occur at the superconducting transition. 
In the crystals in which the vector order parameter (${\boldsymbol d}$ vector) is tilted from the crystal high-symmetry directions as evidenced by nematic diamagnetic susceptibility, we find a sizable 
 lattice distortion ($\sim$100 ppm) associated with the onset of superconductivity. In contrast, in crystals with the  ${\boldsymbol d}$ vector aligned along  the  high-symmetry directions, we find no appreciable change in lattice constant. %the lattice distortion is %negligibly small.
 %so small that it is beyond detection resolution. 
%In thes e distorted samples, angular-dependent in-plane magnetic susceptibility measurements show that the  ${\boldsymbol d}$ vector tilted from the high-symmetry directions. 
Together with a pronounced vestigial behavior of the distortion,  the results are clear evidence for an  odd-parity $E_u$ order parameter that couples with  trigonal lattice. 
Furthermore, in 
the crystal with $x$ = 0.46 where diamagnetic susceptibility is isotropic in the plane,
no lattice distortion accompanying the superconducting transition is found, which  is in line with  a chiral superconducting state %stemming from the two-component order parameter 
in the highly doped region. Our work shows that lattice distortion can be  a powerful diagnosing quantity for nematic superconductivity with two-component order parameter.

%	The relationship between superconductivity and crystal structure has been studied for a long time,
%	but the relationship between superconducting transitions and crystals has hardly been discussed.
%	In this study, We measured the temperature dependence of the lattice constant near the superconducting transition temperature $T_{\rm c}$ of topological superconductor Cu$_x$Bi$_2$Se$_3$ with high precision using synchrotron radiation.
%	We discovered that the slope of the temperature dependence in the lattice constant varies around $T_{\rm c}$.
%	Furthermore, when we applied a magnetic field to suppress superconductivity, the changes around $T_{\rm c}$ were absent. 
%	These results suggest that a minor crystal distortion occurs with the superconducting transition, which can only be explained by triplet superconductivity.
\end{abstract}

\maketitle
%\section{Introduction}
%The relationship between superconductivity and crystal structure has long been debated.  
%In systems lacking inversion symmetry, their connection to spin-triplet superconductivity has also been actively discussed\cite{Gorkov2001,Frigeri2004,Frigeri2004a,Yuan2006,Nishiyama2007,harada2012}.
Whether a superconducting transition accompanies a crystal structural change    has been a topic of research for long time in spin-singlet superconductors.   %with unconventional superconducting nature.
 A volume expansion at the  transition into an isotropic superconducting state was  discussed already in the early stages of superconductivity research \cite{Rohner1957}.  Later, 
a spontaneous lattice distortion at the superconducting transition was predicted for heavy Fermion compounds with an anisotropic gap function \cite{Joynt_Rice1985,Ozaki1986}. It was pointed out that the lattice can expand in the directions along which the gap is large \cite{Joynt_Rice1985}.
 Lattice distortion was also discussed theoretically
in cuprate high-temperature superconductors \cite{Millis}, and a volume expansion associated with the superconducting transition was observed experimentally, hinting at a role of phonon coupling \cite{You1991_cupdate_distortion}. 
%providing evidence of unconventional superconductivity. 
%However, little is known about a spin-triplet  superconductor.

%Notably, the spin-triplet topological superconductor serves as a novel platform for investigating the interactions between superconducting transitions and crystal lattice dynamics.
%Following the discovery of topological insulators, which possess a bulk energy gap while hosting gapless conductive states at their boundaries\cite{Fu2007,Xia2009,Zhang2009,Hsieh2009}
%it was soon proposed that analogous topological superconductors could also exist\cite{Fu2008,Schnyder2008,Qi2009,Tanaka2011,Sasaki2015,Mizushima2016}.  
%This type of superconductor has a nodeless superconducting gap in the bulk while exhibiting gapless surface states at the boundary,
%whose quasiparticle excitations may upport Majorana zero-energy modes\cite{Ivanov2001,Qi_RevModPhys.83.1057}.
%Majorana quasiparticles are believed to obey non-Abelian statistics and are anticipated to have applications in fault-tolerant quantum computing
%\cite{Wilczek2009,Akhmerov2009,Leijnse2012}.

Recently, efforts have been made to turn the  
 topological insulator {\BS}, which has a trigonal primitive lattice belonging to $D_{3d}$ point group, into a superconductor by doping carriers %have attracted attention as platforms that retain topological characteristics\cite{Wray2010} while exhibiting superconductivity
\cite{Hor2010,Qiu_arXiv:1512.03519,Liu2015}.  
In the case of {\cx}, nuclear magnetic resonance  measurements revealed a broken spin-rotation symmetry  below {\tc} \cite{Matano2016},  
%leading to a twofold symmetry that is lower than the crystal's inherent trigonal symmetry, 
which is a hallmark evidence for spin-triplet superconductivity.  
Since the inversion symmetry is preserved in this system, the spin-triplet pairing means an odd-parity  orbital wave function, % of the superconductivity, 
thereby  establishing  {\cx} as a topological superconductor \cite{FuPRL,Sato}, which can host gapless conductive states at their boundaries. %\cite{Fu2007,Xia2009,Zhang2009,Hsieh2009}. 
%it was soon proposed that analogous topological superconductors could also exist\cite{Fu2008,Schnyder2008,Qi2009,Tanaka2011,Sasaki2015,Mizushima2016}.  
%This type of superconductor has a nodeless superconducting gap in the bulk while exhibiting gapless surface states at the boundary,
In particular,  quasiparticle excitations, namely, the Majorana zero-energy modes can exist in the vortex core \cite{Ivanov2001,Qi_RevModPhys.83.1057}, 
which  obey non-Abelian statistics and are expected to be applied in fault-tolerant quantum computing
\cite{Kitaev,Leijnse2012}.
%A noteworthy feature of this system is that the  ${\boldsymbol d}$ vector, representing spin-triplet superconductivity, is pinned to a particular crystallographic axis.
%This suggests that the  ${\boldsymbol d}$vector, associated with spin-triplet pairing, is pinned to a specific crystallographic axis,  
%thus giving rise to a nematic superconducting state\cite{Matano2016,Yonezawa2019_review,Kawai2020,Yokoyama2023}.  
%Various experiments, including magnetic susceptibility, transport, and specific heat measurements, have confirmed this symmetry breaking
%\cite{Pan2016,Yonezawa2017,Asaba2017,Willa2018,Sun2019,Kawai2020,Yokoyama2023},  
%raising questions about why the  ${\boldsymbol d}$-vector is pinned in a particular direction and how crystal strain influences this orientation selection.
%Study on this question has been conducted in the past.
%At room temperature, synchrotron measurements detected subtle crystal distortions
%\cite{Kuntsevich2018,Kuntsevich2019,Froehlich2020}.  
%It has also been reported that the application of uniaxial pressure can align domains,  
%implying a relationship between strain and the direction of the  ${\boldsymbol d}$-vector\cite{Kostylev2020}.  

The discovery of spin rotation symmetry breaking in  {\cx} %at $T_c\sim$ 4 K 
makes this system  an excellent platform for studying the physics of spin-triplet superconductivity, which is much less known compared to spin-singlet cases.
It was proposed that the novel state is described by the  $E_u$ representation with  a two-component order parameter \cite{Fu-PRB}. In such case, the order parameter can couple to the trigonal lattice where the in-plane and out-of-plane shear strain tensors transform as the same irreducible  representation  \cite{Hecker_Fernandes2022}.  
Therefore, obtaining experimental evidence for such coupling is crucial for pining down the order parameter in this class of novel superconductors.
Previously, it was found that the in-plane nematic %response of the 
Meissner diamagnetism has a symmetry tiled from the crystal high-symmetry directions, which is   restored by applying a strong enough magnetic field  \cite{Yokoyama2023}. The tilting   has been interpreted as a consequence of phonon coupling % to the superconducting order parameter 
\cite{Yokoyama2023}, but direct evidence is lacking. 
%
%In such a model, the crystal would distort at the superconducting transitio.
%Concerning the pinning of the  ${\boldsymbol d}$-vector, it has been suggested that delicate coupling among the superconducting order parameter, phonons, or crystal distortions may be involvedn.  
Thermal expansion measurements on Nb$_x$Bi$_2$Se$_3$ suggest a minute ($\sim$0.1 ppm) $a$-axis change upon the superconducting transition\cite{Cho2020}, while high-resolution x-ray measurements 
 found no  distortion  in {\sx} \cite{Smylie2024}. Thus, the coupling between the spin-triplet %superconducting 
 order parameter and lattice remains unclear, % for this spin-triplet superconductor
  and   further investigation is needed.

In this Letter, we report on  x-Ray diffraction measurements using high-resolution synchrotron light for electrochemically Cu-doped {\cx} characterized by  angular-dependent in-plane magnetic susceptibility measurements.
%The in-plane magnetic susceptibility measurements yielded twofold symmetric results, as previously observed.
%By measuring crystals with different Cu contents and different  ${\boldsymbol d}$-vector directions, 
We find a spontaneous lattice distortion of the order 100 ppm in crystals with the  ${\boldsymbol d}$ vector tilted away from  the crystal high-symmetry   directions. The distortion starts at a temperature $T^*$ above $T_c$ with  $T^* \sim $ 1.1 $T_c$. However,   
 no appreciable distortion was detected in  crystals 
 where the  ${\boldsymbol d}$ vector aligns along the high-symmetry (axis)  directions. %The size of the lattice distortion is proportional to the tilting angle of the   ${\boldsymbol d}$-vector. 
 Furthermore, for large doping rate ($x$=0.46) where the gap is isotropic in the plane,  no lattice distortion was found either.
  These results reveal the novel feature of the coupling between the   two component $E_u$ type superconducting order parameter and trigonal crystal lattice.
 %not expected for a spin-singlet superconductor with single-component order parameter, and indicate that 
  %that can, and only 
 %couples with   lattices. 
 %
 %More broadly, investigations on two-component spin-triplet superconductivity 
 Our work can also shed light on kagome or iron-based superconductors with two bands and twisted bilayer graphene, where charge-4$e$  state \cite{WangZQ,Fernandes-Fu} or quadruple fermion state \cite{Grinenko} are a focus of study.

%This finding is consistent with theoretical studies considering a spin-triplet superconducting state with a two-component order parameter.
%For samples with large x that do not show twofold symmetry in their susceptibility, 
%we did not observe any lattice distortion at the superconducting transition, 
%reflecting the a chiral superconducting state.

%\section{Methods}
Single crystals of {\cx} were prepared by intercalating Cu into {\BS} by the electrochemical doping method described in Refs. \onlinecite{Kriener2011,Kawai2020}.
First, single crystals of {\BS} were grown by melting stoichiometric mixtures of 
elemental Bi (99.9999\%) and Se (99.999\%) at 850 $^{\rm o}$C for 48 hours in sealed evacuated quartz tubes. 
After melting, the sample was slowly cooled down to 550 $^{\rm o}$C over 48 hours
and kept at the same temperature for 24 hours.
%Elemental starting materials of Bi (99.9999\% purity) and Se (99.999\%) with a stoichiometric ratio were placed in a quarts tube,
%sealed under vacuum, heated in 850$^{\rm o}$C for 48 hours, 
%cooled slowly down to 550 $^{\rm o}$C for 48 hours, and kept at same temperature for 24 hour.
Those melt-grown {\BS} single crystals were cleaved into smaller rectangular pieces of about 14 mg.
They were wound by bare copper wire (dia. 0.05 mm), and used as a working electrode.
A Cu wire with diameter of 0.5 mm  was used both as the counter (CE) and the reference electrode (RE).
We applied a current of 10 $\mu$A in a saturated solution of CuI powder (99.99\%) in acetonitrile (CH$_3$CN).
The obtained crystal samples were then annealed at 560 $^{\rm o}$C for 1 hour in sealed evacuated quartz tubes, and quenched into water.
%After quenching, the samples were covered with epoxy (STYCAST 1266) to  avoid deterioration.
The Cu concentration $x$ was determined from the mass increment of the samples. It is confirmed that the Cu is indeed doped into the crystal by Knight shift measurements which increases with increasing Cu content $x$ \cite{Kawai2020}.
To check the superconducting properties, dc susceptibility measurements were performed using a superconducting quantum interference device (SQUID) with
the vibrating sample magnetometer (VSM).   The angle-dependent in-plane magnetic susceptibility measurements were performed by rotating the sample inside a fixed coil placed in a magnetic field. 
The angle was determined using two Hall sensors \cite{Yokoyama2023}. 
Synchrotron radiation was employed for the measurements of lattice constants with high-resolution. 
These measurements were performed at the BL-3A beamline of Photon Factory
in High Energy Accelerator Research Organization (KEK).
The incident beam was monochromatized by a pair of Si(111) crystals.
%These measurements were performed using 12 keV beam at the BL-3A beamline of Photon Factory in High Energy Accelerator Research Organization(KEK).
%
\begin{figure}[htbp]
	\includegraphics[clip,width=60mm]{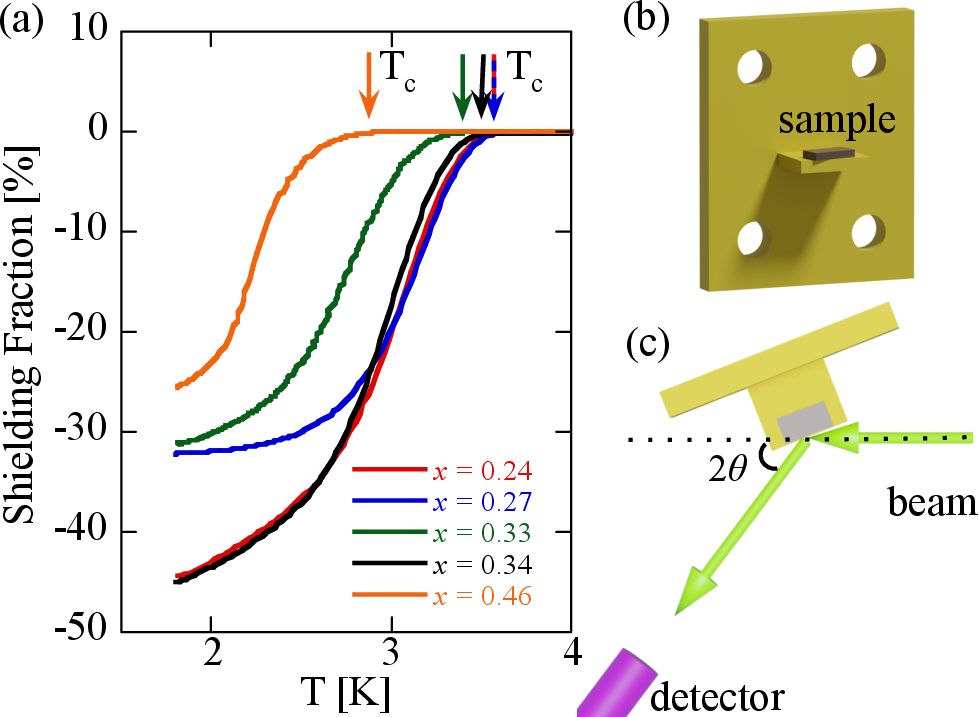}
	\caption{\label{sample1}%color online) 
	(a): Results of the DC magnetization measurements for the crystals used in this study.  
	(b, c): The experimental setup for the synchrotron X-ray diffraction measurements.
	}

\end{figure}
Figure 1(a) shows the results of dc magnetization measurements for the samples used in this study.
%NMR measurements were carried out by using a phase-coherent spectrometer.
%Table 1 summarizes the Cu doping content $x$, the superconducting transition temperature {\tc}, and the superconducting shielding fraction (SF) for  the samples.
Figures 1(b,c) illustrate the experimental setup of the measurements between 1.9 K and 12 K. 
The sample was mounted 
 on a brass holder and measured using a two-axes diffractometer. %, and the beam angle $2\theta$ was varied during data acquisition..
%The temperature range for these measurements was from 1.9 K to 12 K. 
%In the synchrotron radiation experiments,the crystal was fixed, and the beam angle $2\theta$ was varied during data acquisition.
%

%
%
\begin{figure}[htbp]
	\includegraphics[clip,width=70mm]{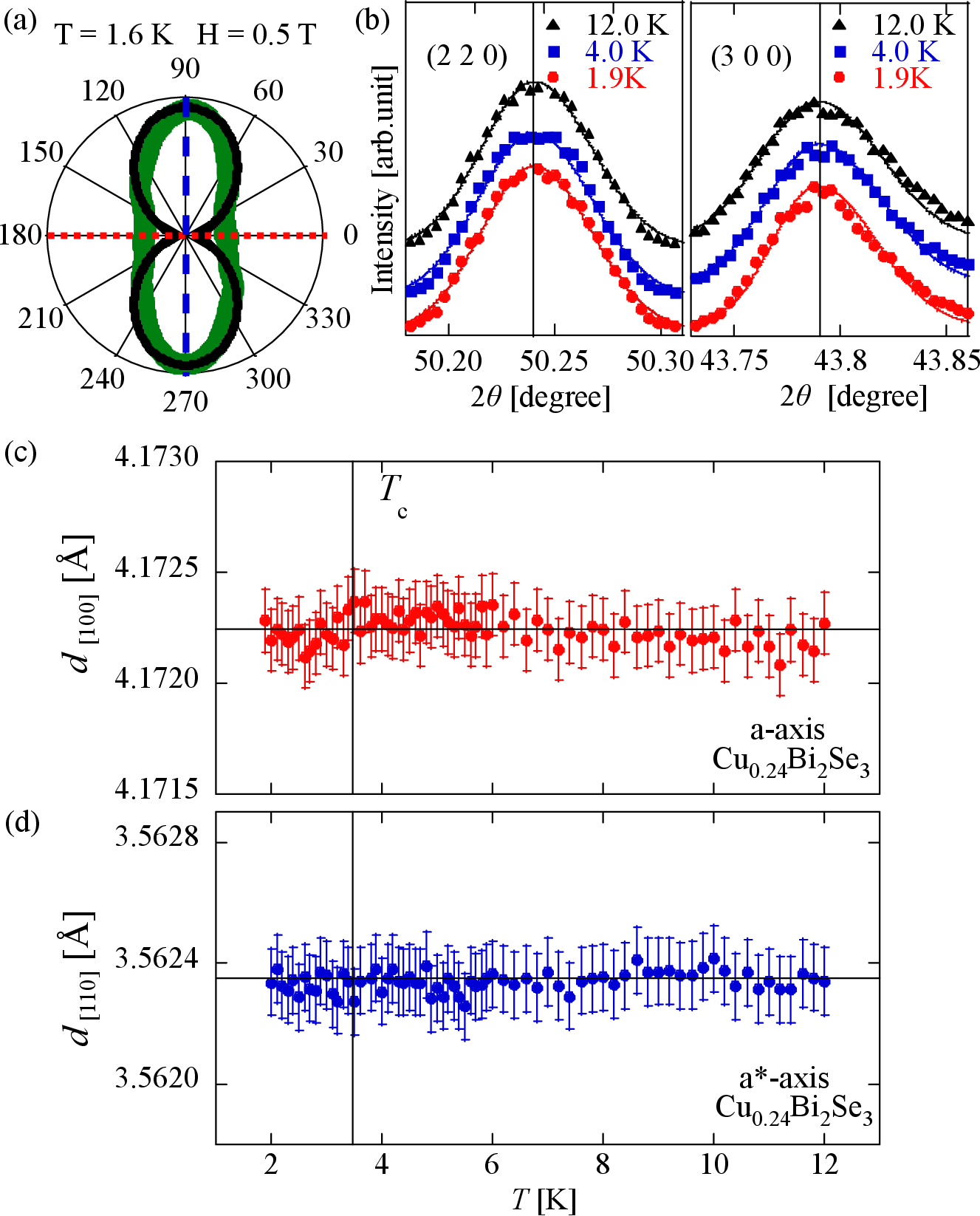}
	\caption{
		\label{sample1}%color online) 
		(a): Diamagnetism measured by  in-plane ac susceptibility as a function of %angle $\phi$ at $T = 1.6 $ K and $H = 0.5 $ T. 
		%The angle is defined with respect to the field direction, and 0 $^{\rm o}$ is parallel to one $a$-axis. 
		%The solid line is a guide to the eye, drawn as $\sin(2\theta+\delta)$, where $\delta$ represents the tilting from the high-symmetry direction.
		the angle $\phi$ between the pre-determined $a$-axis and the applied field. %defined as with respect to the field direction, and 0$^o$ is parallel to one a-axis. 
		The solid black curve is   $\sin(2\phi)$. %, where $\delta$ represents the deviation from the high-symmetry direction (see text).
		The dotted red and blue line corresponds to the direction  along which synchrotron diffraction measurements were performed ( [1 0 0] and [1 1 0]), respectively. % to obtain the $a$-axis lattice constant,
		%while the dashed blue line corresponds to the  (3 0 0) reflection %[1 1 0] % $a^*$-axis
		 % direction (or the (3 0 0) reflection).
		(b): X-ray diffraction spectrum. % of the (2 2 0) and (3 0 0) reflections. 
		%The vertical solid line indicates the peak position above {\tc}. 
		The solid curves are fits  to  two-Gaussian functions.
		(c,d): The lattice-plane distance $d_{[1 0 0]}$ and $d_{[1 1 0]}$ %constant of $a$- and $a^*$-axis 
		calculated from the peak positions  at the (2 2 0) and (3 0 0) reflections, respectively.. The $d_{[1 0 0]}$ is in agreement with the $a$-axis constant reported previously \cite{Hor2010}. 
		Error bars represent the uncertainties from Gaussian fits.}
\end{figure}

\begin{figure}[htbp]
	\includegraphics[clip,width=70mm]{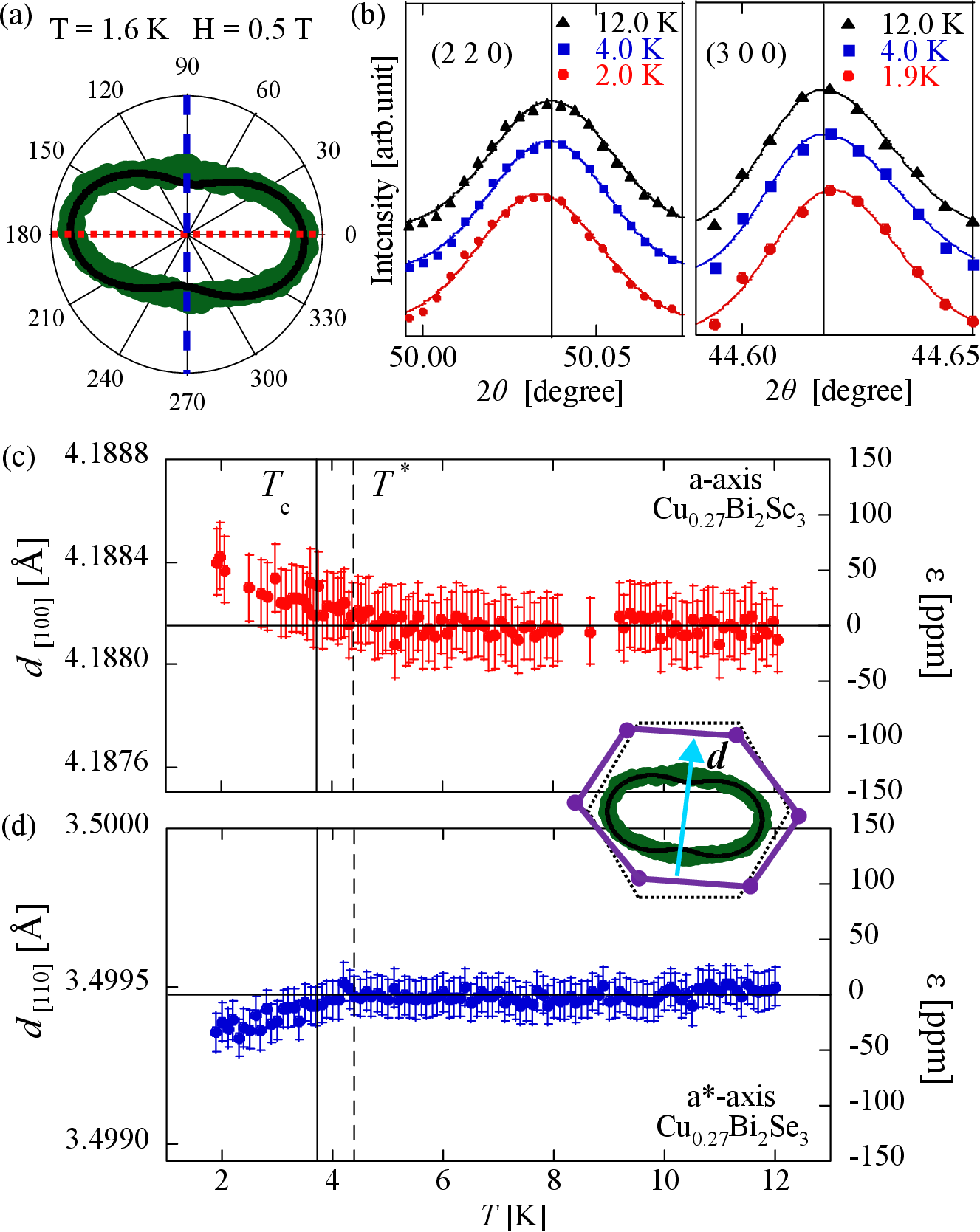}
	\caption{
		\label{conclusion}%color online) 
		(a): %The in-plane ac susceptibility as a function of 
		Angle dependence of the
		diamagnetism. The solid curve is   $\sin(2\phi+\delta)$, where $\delta$ represents the deviation from the high-symmetry direction. % at $T = 1.6 $ K and $H = 0.5 $ T.
	%	The dotted red line corresponds to the direction along which synchrotron diffraction　measurements were performed (the (2 2 0) reflection), % to obtain the $a$-axis lattice constant,
	%	while the dashed blue line corresponds to the (3 0 0) reflection.
		(b): Synchrotron X-ray diffraction spectral. % of the (2 2 0) and (3 0 0) reflections. 
		%The vertical solid line indicates the peak position above {\tc}. 
		%The solid lines are fits to  Gaussian functions.
		(c,d): The lattice-plane distance $d_{[1 0 0]}$ and $d_{[1 1 0]}$ %The lattice constant of $a$ and $a^*$-axis 
		calculated from the peak positions. 
		$\epsilon$ is the deviation rate from the value at temperatures well above {\tc} (see text).
		The inset shows a schematic illustration of the in-plane crystal distortion. The arrow indicates the  ${\boldsymbol d}$ vector direction.
	}
\end{figure}
First, we present the results for $x$ = 0.24.  
Figure 2(a) shows the in-plane ac susceptibility as a function of the field angle at $T$ = 1.6 K and $H$ = 0.5 T in the superconducting state \cite{Yokoyama2023}. As in the previous reports, 
a twofold symmetry with an ellipse shape is observed, in which the upper critical field $H_{c2}$ is the largest along the long axis of the ellipse.
 The  ${\boldsymbol d}$ vector is perpendicular to the long axis of the ellipse \cite{Matano2016,Kawai2020,Yokoyama2023}, and thus is  along the $a$-axis for this crystal. %, with no tilt ($\delta = 0^{\rm o}$).  
 Such nematicity is robust against thermal cycling. 
As found previously \cite{Kawai2020,Yokoyama2023} and also in this work \cite{SM}, 
 regardless  of $x$, {\cx} ($x<$0.4) can host two kinds of nematic superconducting states with the  ${\boldsymbol d}$ vector either parallel or perpendicular to the $a$-axis,   
%one can make two kinds of superconducting {\cx} with the  ${\boldsymbol d}$-vector parallel or perpendicular to the $a$-axis, 
which is believed to correspond to the proposed two (almost) degenerate gap functions of $E_u$ \cite{Fu-PRB}.
Figure 2(b) displays the synchrotron diffraction spectrum of %$\theta$-2$\theta$ scan of %measured along the % $a$-axis 
(2 2 0) and 
%$a^*$-axis 
(3 0 0) reflections. The index (2 2 0) and (3 0 0) correspond to [1 0 0] and [1 1 0] crystal directions. %directions, respectively.
There is no detectable change across {\tc}. 
The temperature dependence of the lattice-plane distance $d_{[1 0 0]}$ %(lattice constant $a$) 
and $d_{[1 1 0]}$  determined from the peak position is shown in Fig. 2(c,d), and no difference is seen above and below {\tc}.
 %constant above {\ts}.{\ts} is the temperature at which nematic fluctuations start and will be discussed in detail later.
%We have also measured a similar crystal with $x$=0.34 and the same result was obtained \cite{SM}.% (see Supplemental Materials \cite{SM}). 
%These results ares consistent with a previous report \cite{Smylie2024}.  

By contrast, we discover a quite large lattice change in crystals with the  ${\boldsymbol d}$ vector tilted away from the high-symmetry direction. %crystal axis. 
Figure 3(a) shows the result for $x$ = 0.27.
The in-plane magnetic susceptibility exhibited a twofold symmetry, but the long axis of the ellipse is tilted away  from the $a^*$-axis by 5$^{\rm o}$. 
%In the sample with $x$ = 0.24, the  ${\boldsymbol d}$-vector was oriented along the $a^*$ direction, but in the sample with $x$ = 0.27, it was tilted by 5$^{\rm o}$ from the $a^*$-axis.
Synchrotron diffraction spectra revealed a spectral  shift below $\sim${\tc}, and  
 the directions of the shift is opposite between the [1 0 0] 
 and [1 1 0] direction. %$a$- and $a^*$-axis.
 The temperature dependence of the lattice-plane distance  is shown in Fig.3(c,d).
The lattice expands along the $a$-axis direction %(by 70 ppm) 
and contracts along the $a^*$-axis direction (Fig. 3 (c,d) inset).
We define $\epsilon = \frac{d(T)-d_0}{d_0}$,  %\times 100$, 
the deviation rate of the lattice-plane distance  from the value at temperatures well above {\tc}, where 
$d(T)$ represents the value at a  temperature $T$, while $d_0$  at $T$=12 K. 
$\epsilon$ is found to be  70 ppm in this crystal.
%We confirmed that such lattice distortion is indeed due to the superconducting transition by applying a magnetic field larger than the upper critical field $H_{c2}$ which eliminates the change seen at zero field (see Supplemental Materials \cite{SM}).

Below we discuss the origin of the lattice distortion below the superconducting transition. 
The total free energy of the system at zero magnetic field consists of contributions from the spin-triplet superconducting order which is nematic,   $S_{\rm triplet}$,  the acoustic phonons (the fluctuations of strain) energy $S_{\rm phonon}$, 
 the coupling between the two, $S_{\rm triplet-phonon}$, and the pinning energy $S_{\rm pinning}$ to lock the ${\boldsymbol d}$ vector to a particular direction:
\begin{equation}
	S_{\rm total} = S_{\rm triplet} +S_{\rm pinning} + S_{\rm phonon} + S_{\rm triplet\text{-}phonon}
\end{equation}
The coupling term is given by \cite{Hecker_Fernandes2022}:
\begin{equation}
		S_{\rm triplet\text{-}phonon} = \int \left\lbrace  {\boldsymbol \Psi} \cdot \left( \kappa_1 {\boldsymbol e}_{1} + \kappa_2 {\boldsymbol e}_{2} \right) \right\rbrace
	%S_{\rm triplet\text{-}phonon} = \int_x \left\lbrace  {\boldsymbol \Psi^{E_g}} \cdot \left( \kappa_1 {\boldsymbol \epsilon}^{E_g,1} + \kappa_2 {\boldsymbol \epsilon}^{E_g,2} \right) \right\rbrace
\end{equation}
where  ${\boldsymbol \Psi}$ = ($\Psi_{1}, \Psi_{2}$) % ($\Psi^{E_g,1}, \Psi^{E_g,2}$) %i$\Psi\sum_{\alpha\beta}\langle \beta \lvert {\boldsymbol d}\cdot {\boldsymbol \sigma} \rvert\sigma_y|\alpha \rangle$ %=(\Psi_x, \Psi_y)
=(${\boldsymbol \Delta}^* {\boldsymbol \sigma}^z {\boldsymbol\Delta}, - {\boldsymbol \Delta}^*  {\boldsymbol \sigma^x}  {\boldsymbol \Delta }$), with ${\boldsymbol \Delta}$ being the two-component spin-triplet nematic order parameter \cite{Fu-PRB} and ${\boldsymbol \sigma}$ the Pauli matrix. 
%, ${\boldsymbol \Psi}$ is the two-component spin-triplet nematic order ${\boldsymbol \Psi}=(\Psi_x, \Psi_y)$ and
 ${\boldsymbol e}_{1,2}$ is the in-plane shear
strain and out-of-plane shear
strain doublet, respectively. 
% ${\boldsymbol \epsilon}^{1}$=
%\begin{pmatrix}
%	\epsilon_{11} - \epsilon_{22}\\
%	-2\epsilon_{12} \\
%\end{pmatrix}
\[
\boldsymbol e_{1}
=
\left(
\begin{array}{r}
	e_{xx} - e_{yy}\\
-2e_{xy} \\
\end{array}
\right)
,
%\]
%\[
\boldsymbol e_{2}
=
\left(
\begin{array}{r}
	2e_{yz} \\
	-2e_{zx}
\end{array}
\right)
\]
The  ${\boldsymbol d}$ vector is  ${\boldsymbol d}$=$(\Delta_xk_z, \Delta_yk_z, -\frac{v_0}{v_z}(\Delta_x k_x+\Delta_y k_y) )$ \cite{Hecker2018}.
Both ${\boldsymbol \Psi}$ and 
 ${\boldsymbol e}_{1,2}$  transform as the same %under the  
%${E_g}$ 
irreducible representation \cite{Hecker_Fernandes2022}, so that they can couple % with ${\boldsymbol \Psi}$ = ($\Psi^{E_g,1}, \Psi^{E_g,2}$) 
through the nematoelastic coupling constants
$\kappa_1$ and $\kappa_2$. 
%This is a special property of trigonal primitive lattice not possessed by hexagonal lattices.
This linear coupling mediates the lattice distortion.

The intrinsic spin-triplet %nematic anharmonic cubic
 interaction $S_{\rm triplet}$ favors the  ${\boldsymbol d}$ vector (nematic director) to align parallel to the high-symmetry directions, and the pinning energy $S_{\rm pinning}$ is to select one direction among them \cite{Yokoyama2023}. 
The phonon-mediated %nonanalytic quadratic
interaction prefers the nematic director to align to the directions farthest away from the high-symmetry directions. When the nematoelastic coupling is large enough, it  can unlock the nematic director from the high-symmetry directions. Therefore, the tilting of the  ${\boldsymbol d}$ vector can be understood as due to a large coupling between ${\boldsymbol \Psi}$ and ${\boldsymbol e}_{1,2}$ \cite{Yokoyama2023}, which can vary from crystal to crystal as the defects in the sample is un-controllable owing to the quenching process during crystal synthesis.
When there are more defects in the sample, acoustic (long wave) phonon propagation could be hindered, resulting in a smaller nematoelastic coupling,  whose detailed mechanism needs more investigation in the future though. At any rate, 
%$S_{\rm el}$ is a term originating from the strain in the crystal, and 
%it varies depending on the sample.
%The orientation of the  ${\boldsymbol d}$ vector is determined by $S_{\rm total}$.
%The tilting varies depending on the sample because $S_{\rm el}$ differs for each sample.
%It is considered that the variation in the d-vector orientation across samples is due to this effect.
the direct consequence of a large  nematoelastic coupling is %The same theory also points out that the symmetry of the crystal distortion associated with the superconducting transition varies depending on the orientation of the  ${\boldsymbol d}$ vector. When the  ${\boldsymbol d}$-vector aligns with a high-symmetry direction, the 
a large crystal distortion. % into a triclinic phase \cite{Hecker_Fernandes2022}. %, whereas a tilted  ${\boldsymbol d}$-vector leads to a triclinic distortion.

Although %Next, we discuss why a lattice distortion is observed at the superconducting transition in some cases but not in others. 
both the $x$ = 0.24 and $x$ = 0.27 samples exhibit a twofold symmetry at low fields,  in the former  
 the  ${\boldsymbol d}$ vector is oriented along a high-symmetry axis ($a$-axis) of the crystal, 
whereas in the $x$ = 0.27 sample, the  ${\boldsymbol d}$ vector is tilted away from $a^*$-axis direction, meaning that the nematoelastic coupling is larger in the later. %This trend is seen in all samples we made and measured \cite{SM}.
The previously reported  results for {\sx} and {\nx}  \cite{Smylie2024,Cho2020} can also be well reconciled. The electrical resistivity  measurements show that the $H_{c2}$ shows an ellipse-like two-fold symmetry with the long-axis of the ellipse along  crystal axis. Therefore, the nematoelastic coupling is small there. As a result, the lattice change due to the superconducting transition is expected to be very small, and is indeed in the order of 0.1 ppm as seen in the high-resolution thermal expansion measurement \cite{Cho2020} which was unable to be detected by Smylie {\it et al} \cite{Smylie2024}
and in two of our crystals ($x$=0.24 and 0.33 \cite{SM}) simply because it is beyond the resolution limit of X-ray measurement.
 
In passing, we note two 
important properties of the spontaneous lattice distortion. First, the $c$-axis is also distorted \cite{SM}, therefore the crystal structure is in a triclinic phase under a larger nematoelastic coupling. %, as predicted  theoretically \cite{Hecker_Fernandes2022}.  
Second,  a close inspection of the data shows that the lattice changes occur at a temperature {\ts} slightly above $T_c$.  
Figure 4 plots the  values of {\ts} against {\tc}, including the data for  {\nx} \cite{Cho2020} and {\sx} where specific heat measurement found an anomaly ascribable to superconductivity above a temperature above the  {\tc} \cite{Sun2019}. %determined by resistivity measurement 
A good correlation is seen between {\ts} and {\tc}. This is another piece of evidence for two-component $E_u$ order parameter in {\cx } as elaborated below.
\begin{figure}[htbp]
	\includegraphics[clip,width=60mm]{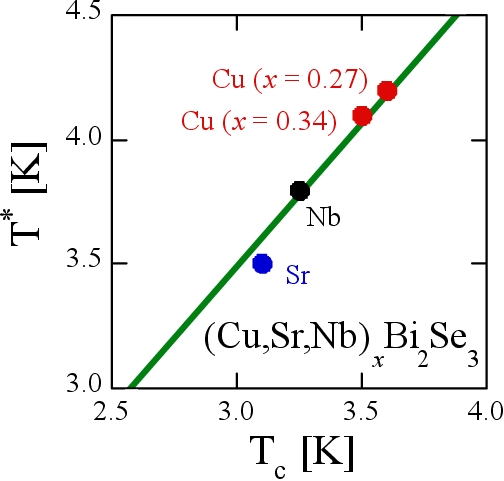}
	\caption{
		\label{diagram}%color online) 
		%	(a): The tilting from the high-symmetry axis of twofold in-plane symmetry $\delta$ against $\epsilon$, the magnitude of the distortion accompanying the superconducting transition.
The precursor temperature  {\ts} correlated with {\tc}. The data point  for Sr is  from the specific measurement \cite{Sun2019}, and the data point for Nb is from the thermal expansion measurement \cite{Cho2020}.	The straight line is a linear fit to $T^*=k T_c$ with the result $k$=1.16.
	%	The solid line is a guide for the eyes. %linear fitting to the data.
	}
\end{figure}

A spin-triplet two-component phase   is
characterized by a 2$\times$2 composite order parameter $\mathbf{Q}$ \cite{Hecker2018}, with the tensor % $	\mathbf{Q}_{\alpha\beta} = - \sum_{\mu\nu}{{\boldsymbol \sigma}_{\mu\nu}{\boldsymbol \sigma}_{\alpha\beta}\Delta_{\mu}^*\Delta_{\nu}}$
$\mathbf{Q}_{\alpha\beta} =  \sum_{\mu\nu}{{\boldsymbol \sigma}_{\mu\nu}{\boldsymbol \sigma}_{\alpha\beta}\Delta_{\mu}^*\Delta_{\nu}}$, where ${\boldsymbol \sigma}=(\sigma_z, \sigma_x)$. Explicitly, $\mathbf{Q}$ can be expressed as 
\begin{equation}
%	\mathbf{Q}_{\alpha\beta} =  \sum_{\mu\nu}{{\boldsymbol \sigma}_{\mu\nu}{\boldsymbol %\sigma}_{\alpha\beta}\Delta_{\mu}^*\Delta_{\nu}}
	\mathbf{Q} =
	\begin{pmatrix}
		|\Delta_x|^2 - |\Delta_y|^2 & \Delta_x^* \Delta_y + \Delta_y^* \Delta_x \\
		\Delta_x^* \Delta_y + \Delta_y^* \Delta_x & |\Delta_y|^2 - |\Delta_x|^2
	\end{pmatrix}.
\end{equation}
The  expectation value  $\langle\Delta_x\rangle$ = $\langle\Delta_y\rangle$ = 0  above $T_c$, while the expectation value of $\mathbf{Q}$ can be finite above $T_c$, since $\langle \Delta_x^*\Delta_y \rangle \neq$ 0 for example. 
As a result, a lattice distortion $\boldsymbol{e}_{\mu\nu}$$\propto \kappa$$\mathbf{Q}_{\mu\nu}$ can develop at a temperature {\ts} above $T_c$.
%The result that  {\ts} is larger than  {\tc} by about  10\%  {\tc}  is consistent with an  estimate  using the elastic constant appropriate for {\cx} \cite{Hecker2018}.
%
%It has further been proposed that fluctuations associated with the two-component superconducting order induce spontaneous lattice distortion even above {\tc}.
%
%In {\cx}, the superconducting order parameter is two-component and can be expressed as $\Delta= \Delta_0 e^{i\phi}(\cos\theta, \sin\theta) = (\Delta_x,\Delta_y)$, where $\Delta_0$ denotes the amplitude, $\phi$ is the global $U(1)$ phase, and $\theta$ is the angle that selects a specific crystal axis.The existence of two order parameters gives rise to a nematic phase above {\tc},
%characterized by a finite expectation value of the composite order parameter:
%
%The fluctuation term drives the onset of crystal distortion at temperatures above {\tc}. 
Therefore, %although paradoxically, 
the large %temperature 
discrepancy between {\tc} and  {\ts} by about 15\% {\tc}, which is not encountered in conventional superconductors where the superconducting-fluctuation temperature range is extremely small and is  not accessible by experiments \cite{Varlamov}, serves as a hallmark for a  two-component $E_u$ superconducting order parameter.
We note that the  phenomena of charge-4$e$ state \cite{WangZQ,Fernandes-Fu} or quartic fermion phase \cite{Grinenko} share similarities with our work in that %correlations between Cooper pairs 
two components of the order parameter play an essential role.

%In contrast, the superconducting phase is defined by finite expectation values of $(\Delta_x$ or $\Delta_y)$.
%Hecker and Fernandes recently proposed that competition between a quadratic phonon-mediated interaction 
%and cubic nematic anharmonicity can cause the nematic director to rotate away from high-symmetry directions\cite{Hecker_Fernandes2022}. 
%The two-component superconducting state couples linearly to phonons, resulting in a rotation of the superconducting symmetry. 
%As {\tc} increases, {\ts} also increases, consistent with the proposal that {\ts} scales with {\tc}\cite{Hecker2018}.
%This deviation from the high-symmetry direction is believed to arise from the coupling between the two-component superconducting order parameter and phonons\cite{Hecker_Fernandes2022}. 
%
\begin{figure}[htbp]
	\includegraphics[clip,width=60mm]{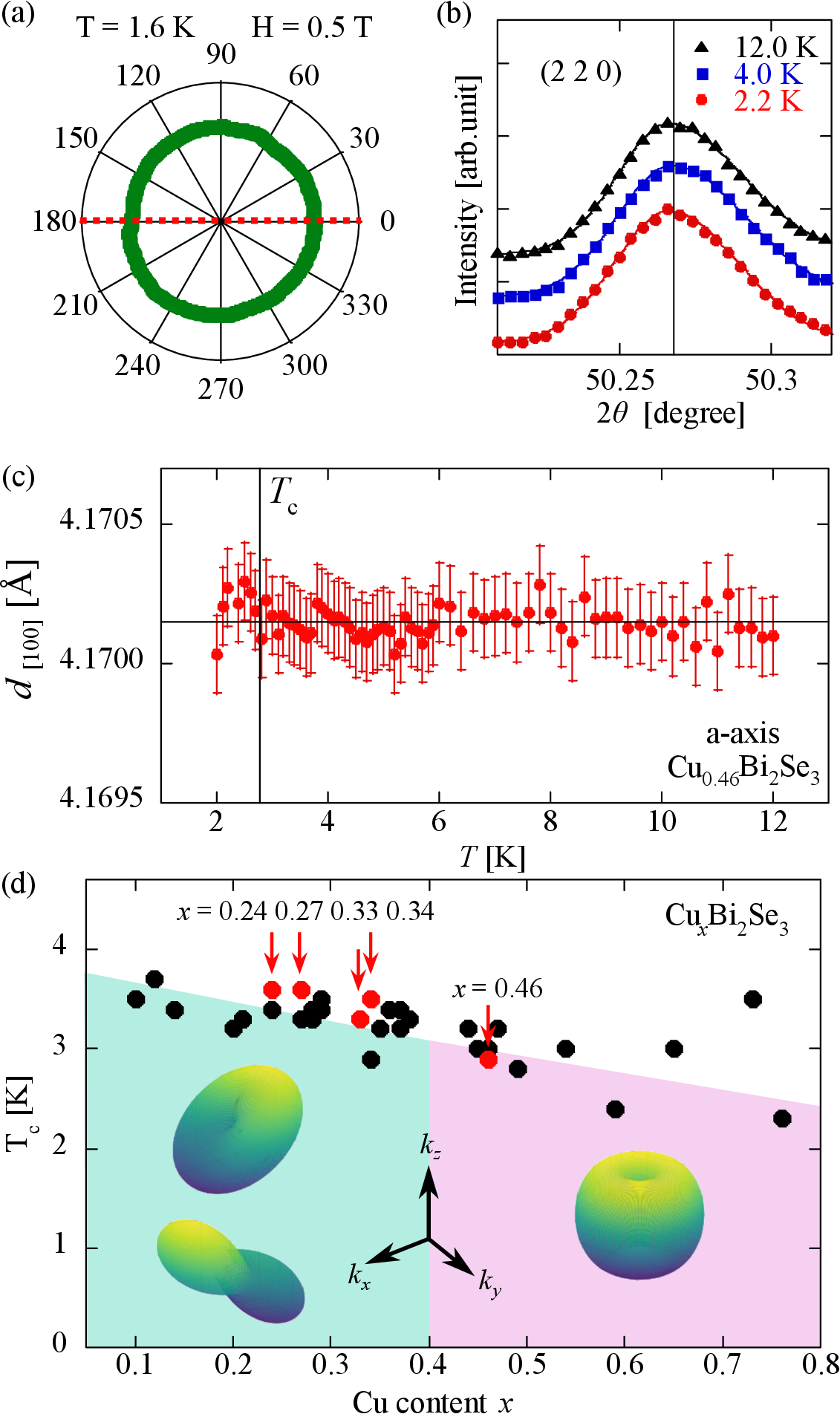}
	\caption{
		\label{sample3}%color online) 
		(a): The in-plane diamagnetism for $x$ = 0.46. % at %ac susceptibility as a function of angle in 
		%	$T = 1.6 $ K and $H = 0.5 $ T.
		%	The dotted red line corresponds to the axis along which synchrotron　diffraction　measurements were performed ( the (2 2 0) reflection).
		(b): Synchrotron X-ray diffraction spectra of the (2 2 0) refection. 
		%The vertical solid line indicates the peak position above {\tc}. 
		%The solid lines fitted to the data are Gaussian curves.
		(c): The lattice-plane distance $d_{[1 0 0]}$ and $d_{[1 1 0]}$ %The lattice constant of $a$-axis 
		calculated from the peak positions. 
		(d): Cu content $x$ dependence of superconducting critical temperature {\tc}.
		The arrows indicate samples measured in this study. %for which the results are presented in the main text. 
		The illustrations in each region show the gap functions of the two nematic states and one of the chiral states \cite{Fu-PRB}, respectively. % See text for more detail. %For the chiral states, the left image shows the nodal gap and the right image shows the full gap.
	} %($\Delta_x$,  $\Delta_y$ and $\Delta_x+i\Delta_y$) 
\end{figure}
Finally, Fig. 5 shows the results for $x$ = 0.46.  
Unlike the results for $x<$0.4, the angular dependence of the in-plane magnetic susceptibility yielded an isotropic shape.
In synchrotron x-ray diffraction measurements, no changes were observed in the spectra around {\tc}. 
Namely, the lattice constant exhibited no significant change across {\tc}. We should first comment on a possible extrinsic cause for the lack of lattice distortion. 
In the high-doping regime, one may suspect that multiple domains are present which  may cause an isotropic in-plane susceptibility and nulls any lattice change. 
In our synchrotron measurements, the beam size was 0.05 $\times$ 0.05 mm$^{2}$. 
If domains smaller than beam size were present, distortions occurring in different directions would broaden the spectral peaks upon the superconducting transition.
However, no such broadening was observed. Therefore we believe that the observed result is an  intrinsic property.

Our previous research has clarified that the symmetry of the superconducting gap changes from nematic to isotropic at around $x$ = 0.4\cite{Kawai2020}. 
In angular-dependent measurements of the in-plane upper critical field {\hct}, the low-doping regime shows twofold symmetry, 
while the high-doping regime exhibits isotropic behavior. 
It has been pointed out that \cite{Fu-PRB}, the two-component order parameter can result  in either a nematic state $\Delta_x$ or $\Delta_y$, or a chiral state % $\Delta_x$ + $i \Delta_y$ 
in which the gap is isotropic in the hexagonal plane. The gap of the chiral state depends on spin species, one having a full gap and the other nodal gap. % 
%Although there are
  %of the $\Delta_x$ + $i \Delta_y$ state 
% (see Fig.5(c)). 
 Even the later one with two nodes at the poles can be  energetically stable %than a nematic one 
 as the Fermi surface becomes more two-dimensional at higher doping \cite{Arpes}.
%A theoretical model has been proposed in which the Fermi surface becomes more two-dimensional at higher doping, resulting in a dominant chiral state.
%For the $x$ = 0.46 sample, the in-plane ac suceptibility show isotropic behavior, consistent with previous reports.
The synchrotron x-ray diffraction experiments revealing no in-plane distortion accompanying the superconducting transition is  consistent with such chiral superconducting state.
Therefore, our new results further solidify the phase diagram schematically shown in Fig. 5, providing additional supporting evidence for the doping-dependent changes in the pairing symmetry.

%\section{Conclusion}
In summary, we performed synchrotron X-ray diffraction measurements to investigate whether a lattice distortion occurs at the superconducting transition of the spin-triplet topological superconductor {\cx}. 
The crystals in which the  ${\boldsymbol d}$ vector is tilted from the crystal high-symmetry directions as evidenced by nematic diamagnetic susceptibility, we find a sizable   
lattice distortion ($\sim$100 ppm) associated with the onset of superconductivity. % is so small that it is beyond the detection resolution. %The amount of the distortion is proportional to the  ${\boldsymbol d}$-vector deviation angle, and 
The distortion shows a 
%In these distorted samples, angular-dependent in-plane magnetic susceptibility measurements show that the  ${\boldsymbol d}$ vector tilted from the high-symmetry directions. 
 pronounced vestigial behavior. 
  In contrast, in crystals with the  ${\boldsymbol d}$ vector aligned along  the  high-symmetry directions, we find no appreciable  lattice distortion.
  These results are clear evidence for a  two-component order parameter that couples with  trigonal lattice. 
Furthermore, 
the crystal with $x$ = 0.46 where the gap  is isotropic in the plane,
no lattice distortion accompanying the superconducting transition is detected, which  is in line with  a chiral superconducting state in the highly doped region. Our results provide fresh new insights into the physics of superconductivity with two-component odd-parity order parameters. %Importantly, our work show that lattice distortion can be used as a powerful probe  for identifying nematic superconductivity with two-components order parameter.
%Among several samples studied, we discovered a lattice distortion emerging at the superconducting transition. The sample that showed such distortion also exhibited a  ${\boldsymbol d}$-vector tilted away from the high-symmetry directions in the in-plane angular dependence of the ac susceptibility. On the other hand, samples whose  ${\boldsymbol d}$-vectors aligned along high-symmetry directions displayed no lattice distortion. Furthermore, for the $x$ = 0.46 sample, where the in-plane angular dependence is isotropic, no lattice distortion was observed at the superconducting transition. 
%This finding suggests that, in the high-doping regime, a chiral superconducting state is realized and does not lead to lattice distortion.

%\begin{acknowledgments}
We thank R. Fernandes and M. Hecker for stimulating discussions that lead to initiating this work. We are particularly grateful to K. Ishii  for coordinating the measurements in KEK.  We also thank N. Ikeda and Y. Nogami for helpful discussion on synchrotron light experiments, and A. Kobayashi and S. Kawasaki for participating in the early stage of this project. This work was performed under the approval of the Photon Factory Program Advisory Committee (Proposal No. 2022G657), and supported by KAKENHI (JPSJ grant No. 22H04482 and No. 25K07207) and the Okayama Foundation for Science and Technology.
%\end{acknowledgments}
%\bibliographystyle{apsrev4-1}
%\bibliography{bi2se3,bi2se3_distortion}

%apsrev4-2.bst 2019-01-14 (MD) hand-edited version of apsrev4-1.bst
%Control: key (0)
%Control: author (72) initials jnrlst
%Control: editor formatted (1) identically to author
%Control: production of article title (-1) disabled
%Control: page (0) single
%Control: year (1) truncated
%Control: production of eprint (0) enabled
\begin{thebibliography}{0}%
\makeatletter
\providecommand \@ifxundefined [1]{%
 \@ifx{#1\undefined}
}%
\providecommand \@ifnum [1]{%
 \ifnum #1\expandafter \@firstoftwo
 \else \expandafter \@secondoftwo
 \fi
}%
\providecommand \@ifx [1]{%
 \ifx #1\expandafter \@firstoftwo
 \else \expandafter \@secondoftwo
 \fi
}%
\providecommand \natexlab [1]{#1}%
\providecommand \enquote  [1]{``#1''}%
\providecommand \bibnamefont  [1]{#1}%
\providecommand \bibfnamefont [1]{#1}%
\providecommand \citenamefont [1]{#1}%
\providecommand \href@noop [0]{\@secondoftwo}%
\providecommand \href [0]{\begingroup \@sanitize@url \@href}%
\providecommand \@href[1]{\@@startlink{#1}\@@href}%
\providecommand \@@href[1]{\endgroup#1\@@endlink}%
\providecommand \@sanitize@url [0]{\catcode `\\12\catcode `\$12\catcode
  `\&12\catcode `\#12\catcode `\^12\catcode `\_12\catcode `\%12\relax}%
\providecommand \@@startlink[1]{}%
\providecommand \@@endlink[0]{}%
\providecommand \url  [0]{\begingroup\@sanitize@url \@url }%
\providecommand \@url [1]{\endgroup\@href {#1}{\urlprefix }}%
\providecommand \urlprefix  [0]{URL }%
\providecommand \Eprint [0]{\href }%
\providecommand \doibase [0]{https://doi.org/}%
\providecommand \selectlanguage [0]{\@gobble}%
\providecommand \bibinfo  [0]{\@secondoftwo}%
\providecommand \bibfield  [0]{\@secondoftwo}%
\providecommand \translation [1]{[#1]}%
\providecommand \BibitemOpen [0]{}%
\providecommand \bibitemStop [0]{}%
\providecommand \bibitemNoStop [0]{.\EOS\space}%
\providecommand \EOS [0]{\spacefactor3000\relax}%
\providecommand \BibitemShut  [1]{\csname bibitem#1\endcsname}%
\let\auto@bib@innerbib\@empty
%</preamble>
\end{thebibliography}%


\begin{thebibliography}{10}
%\bibitem{Gorkov2001}
%L. P. Gor’kov and E. I. Rashba, Superconducting 2D System with Lifted Spin Degeneracy:
%Mixed Singlet-Triplet State, Phys. Rev. Lett. {\bf 87}, 037004 (2001).

%\bibitem{Frigeri2004}
%P. A. Frigeri, D. F. Agterberg, and M. Sigrist, Spin susceptibility in superconductors without
%inversion symmetry, New J. Phys. {\bf 6}, 115 (2004).

%\bibitem{Frigeri2004a}
%P. A. Frigeri, D. F. Agterberg, A. Koga, and M. Sigrist, Superconductivity without Inversion Symmetry: MnSi versus CePt$_3$ Si, Phys. Rev. Lett. {\bf 92}, 097001 (2004).

%\bibitem{Yuan2006}
%H. Q. Yuan, D. F. Agterberg, N. Hayashi, P. Badica, D. Vandervelde, K. Togano, M. Sigrist,
%and M. B. Salamon, S-Wave Spin-Triplet Order in Superconductors without Inversion Symmetry: Li$_2$Pd$_3$B and Li$_2$Pt$_3$B, Phys. Rev. Lett. {\bf 97}, 017006 (2006).

%\bibitem{Nishiyama2007}
%M. Nishiyama, Y. Inada, and G.-q. Zheng, Spin Triplet Superconducting State due to Broken
%Inversion Symmetry in Li$_2$Pt$_3$B, Phys. Rev. Lett. {\bf 98}, 047002 (2007).

%\bibitem{harada2012}
%S. Harada, J. J. Zhou, Y. G. Yao, Y. Inada, and G.-q. Zheng, Abrupt enhancement of non-centrosymmetry and appearance of a spin-triplet superconducting state in Li$_2$(Pt$_{1-x}$Pd$_x$)$_3$B13 beyond $x$ = 0.8, Phys. Rev. B {\bf 86}, 220502 (2012).

\bibitem{Rohner1957}
 J. L. Olsen and H. Rohrer, The Volume Change at the Superconducting Transition, Helv. Phys. Acta
{\bf 30}, 49 (1957).

\bibitem{Joynt_Rice1985}
R. Joynt and T. M. Rice, Strain distortion in anisotropic superconductors, Phys. Rev. B {\bf 32},
6074 (1985).

\bibitem{Ozaki1986}
M.-a. Ozaki, Group Theoretical Analysis of the Lattice Distortion in Anisotropic Super-
conductivity, Prog. Theor. Phys. {\bf 76}, 1008 (1986).

\bibitem{Millis}
A. J. Millis and K. M. Rabe, 
Superconductivity and lattice distortions in high-$T_c$ superconductors. Phys. Rev. B {\bf 38}, 8908 (1988).

\bibitem{You1991_cupdate_distortion}
H. You, U. Welp, and Y. Fang, Slope discontinuity and fluctuation of lattice expansion
nearTcin untwinned YBa$_2$Cu$_3$O$_{7-\delta}$ single crystals, Physical Review B {\bf 43}, 3660 (1991).

%\bibitem{Fu2007}
%L. Fu, C. L. Kane, and E. J. Mele, Topological Insulators in Three Dimensions, Phys. Rev.Lett. {\bf 98}, 106803 (2007).

%\bibitem{Xia2009}
%Y. Xia, D. Qian, D. Hsieh, L. Wray, A. Pal, H. Lin, A. Bansil, D. Grauer, Y. S. Hor, R. J.
%Cava, and M. Z. Hasan, Observation of a large-gap topological-insulator class with a single Dirac cone on the surface, Nature Physics {\bf 5}, 398 (2009).

%\bibitem{Zhang2009}
%H. Zhang, C.-X. Liu, X.-L. Qi, X. Dai, Z. Fang, and S.-C. Zhang, Topological insulators in
%Bi$_2$Se$_3$, Bi$_2$Te$_3$ and Sb$_2$Te$_3$ with a single Dirac cone on the surface, Nature Physics {\bf 5}, 438(2009).

%\bibitem{Hsieh2009}
%D. Hsieh, Y. Xia, D. Qian, L. Wray, J. H. Dil, F. Meier, J. Osterwalder, L. Patthey, J. G.Checkelsky, N. P. Ong, A. V. Fedorov, H. Lin, A. Bansil, D. Grauer, Y. S. Hor, R. J. Cava,and M. Z. Hasan, A tunable topological insulator in the spin helical Dirac transport regime,Nature {\bf 460}, 1101 (2009).

%\bibitem{Fu2008}
%L. Fu and C. L. Kane, Superconducting Proximity Effect and Majorana Fermions at the Surface of a Topological Insulator, Phys. Rev. Lett. {\bf 100}, 096407 (2008).

%\bibitem{Schnyder2008}
%A. P. Schnyder, S. Ryu, A. Furusaki, and A. W. W. Ludwig, Classification of topological insulators and superconductors in three spatial dimensions, Phys. Rev. B {\bf 78}, 195125 (2008).

%\bibitem{Qi2009}
%X.-L. Qi, T. L. Hughes, S. Raghu, and S.-C. Zhang, Time-Reversal-Invariant TopologicalSuperconductors and Superfluids in Two and Three Dimensions, Phys. Rev. Lett. {\bf 102}, 187001(2009).

%\bibitem{Tanaka2011}
%Y. Tanaka, M. Sato, and N. Nagaosa, Symmetry and Topology in Superconductors -Odd Frequency Pairing and Edge States-, J. Phys. Soc. Jpn. {\bf 81}, 011013 (2011).14

%\bibitem{Sasaki2015}
%S. Sasaki and T. Mizushima, Superconducting doped topological materials, Physica C: Super-conductivity and its Applications {\bf 514}, 206 (2015).

%\bibitem{Mizushima2016}
%T. Mizushima, Y. Tsutsumi, T. Kawakami, M. Sato, M. Ichioka, and K. Machida, Symmetry-
%Protected Topological Superfluids and Superconductors –From the Basics to 3He–, J. Phys.
%Soc. Jpn. {\bf 85}, 022001 (2016).






%\bibitem{Wray2010}
%L. A. Wray, S.-Y. Xu, Y. Xia, Y. S. Hor, D. Qian, A. V. Fedorov, H. Lin, A. Bansil, R. J. Cava,and M. Z. Hasan, Observation of topological order in a superconducting doped topologicalinsulator, Nature Physics {\bf 6}, 855 (2010).

\bibitem{Hor2010}
Y. S. Hor, A. J. Williams, J. G. Checkelsky, P. Roushan, J. Seo, Q. Xu, H. W. Zandbergen,
A. Yazdani, N. P. Ong, and R. J. Cava, Superconductivity in Cu x Bi 2 Se 3 and its Implications
for Pairing in the Undoped Topological Insulator, Phys. Rev. Lett. {\bf 104}, 057001 (2010).

\bibitem{Qiu_arXiv:1512.03519}
Y. Qiu, K. N. Sanders, J. Dai, J. E. Medvedeva, W. Wu, P. Ghaemi, T. Vojta, and Y. S.
Hor, Time reversal symmetry breaking superconductivity in topological materials (2015),
arXiv:1512.03519 [cond-mat.supr-con].

\bibitem{Liu2015}
Z. Liu, X. Yao, J. Shao, M. Zuo, L. Pi, S. Tan, C. Zhang, and Y. Zhang, Superconductivity
with Topological Surface State in Sr$_x$Bi$_2$Se$_3$ , J. Am. Chem. Soc. {\bf 137}, 10512 (2015).

\bibitem{Matano2016}
K. Matano, M. Kriener, K. Segawa, Y. Ando, and G.-q. Zheng, Spin-rotation symmetry break-
ing in the superconducting state of Cu$_x$Bi$_2$Se$_3$ , Nature Physics {\bf 12}, 852 (2016).

\bibitem{FuPRL}
L. Fu,  and E. Berg,  Odd-parity topological superconductors: theory and
application to CuxBi2Se3. Phys. Rev. Lett. {\bf 105}, 097001 (2010).

\bibitem{Sato}
M. Sato,  Topological odd-parity superconductors. Phys. Rev. B {\bf 81}, 220504 (2010).


\bibitem{Ivanov2001}
D. A. Ivanov, Non-abelian statistics of half-quantum vortices in p-wave superconductors, Phys.
Rev. Lett. {\bf 86}, 268 (2001).

\bibitem{Qi_RevModPhys.83.1057}
X.-L. Qi and S.-C. Zhang, Topological insulators and superconductors, Rev. Mod. Phys. {\bf 83},
1057 (2011).

%\bibitem{Wilczek2009}
%F. Wilczek, Majorana returns, Nature Physics {\bf 5}, 614 (2009).

%\bibitem{Akhmerov2009}
%A. R. Akhmerov, J. Nilsson, and C. W. J. Beenakker, Electrically Detected Interferometry ofMajorana Fermions in a Topological Insulator, Phys. Rev. Lett. {\bf 102}, 216404 (2009).

\bibitem{Kitaev}
A. Y. Kitaev, Unpaired Majorana fermions in quantum wires. Physics-Uspekhi {\bf 44}, 131 
(2001).

\bibitem{Leijnse2012}
M. Leijnse and K. Flensberg, Introduction to topological superconductivity and Majorana fermions, Semiconductor Science and Technology {\bf 27}, 124003 (2012).

\bibitem{Fu-PRB}
J. F. Venderbos, V. Kozii, and L. Fu,  Odd-parity superconductors with two-component order parameters:
Nematic and chiral, full gap, and Majorana node, 
PHYSICAL REVIEW B {\bf 94},180504(R) (2016).

%\bibitem{Pan2016}
%Y. Pan, A. M. Nikitin, G. K. Araizi, Y. K. Huang, Y. Matsushita, T. Naka, and A. de Visser,Rotational symmetry breaking in the topological superconductor Sr$_x$Bi$_2$Se$_3$ probed by upper-critical field experiments, Scientific Reports {\bf 6}, 28632 (2016).

%\bibitem{Yonezawa2017}
%S. Yonezawa, K. Tajiri, S. Nakata, Y. Nagai, Z. Wang, K. Segawa, Y. Ando, and Y. Maeno,Thermodynamic evidence for nematic superconductivity in cuxbi2se3, Nature Physics {\bf 13}, 123(2017).

%\bibitem{Asaba2017}
%T. Asaba, B. J. Lawson, C. Tinsman, L. Chen, P. Corbae, G. Li, Y. Qiu, Y. S. Hor, L. Fu, and
%L. Li, Rotational Symmetry Breaking in a Trigonal Superconductor Nb-doped Bi$_2$Se$_3$ , Phys.Rev. X {\bf 7}, 011009 (2017).

%\bibitem{Willa2018}
%K. Willa, R. Willa, K. W. Song, G. D. Gu, J. A. Schneeloch, R. Zhong, A. E. Koshelev, W.-K.
%Kwok, and U. Welp, Nanocalorimetric evidence for nematic superconductivity in the doped
%topological insulator Sr$_{0.1}$Bi$_2$Se$_3$ , Phys. Rev. B {\bf 98}, 184509 (2018).







\bibitem{Hecker_Fernandes2022}
M. Hecker and R. M. Fernandes, Phonon-induced rotation of the electronic nematic director
in superconducting Bi$_2$Se$_3$ , Phys. Rev. B {\bf 105}, 174504 (2022).

\bibitem{Yokoyama2023}
M. Yokoyama, H. Nishigaki, S. Ogawa, S. Nita, H. Shiokawa, K. Matano, and G.-q. Zheng,
Manipulating the nematic director by magnetic fields in the spin-triplet superconducting state
of Cu$_x$Bi$_2$Se$_3$ , Phys. Rev. B {\bf 107}, L100505 (2023).

\bibitem{Kawai2020}
T. Kawai, C. G. Wang, Y. Kandori, Y. Honoki, K. Matano, T. Kambe, and G.-q. Zheng, Di-
rection and symmetry transition of the vector order parameter in topological superconductors
Cu$_x$Bi$_2$Se$_3$ , Nature Communications {\bf 11}, 235 (2020).

\bibitem{Cho2020}
C.-w. Cho, J. Shen, J. Lyu, O. Atanov, Q. Chen, S. H. Lee, Y. S. Hor, D. J. Gawryluk,
E. Pomjakushina, M. Bartkowiak, M. Hecker, J. Schmalian, and R. Lortz, Z3-vestigial nematic
order due to superconducting fluctuations in the doped topological insulators Nb$_x$Bi$_2$Se$_3$ and
Cu$_x$Bi$_2$Se$_3$ , Nature Communications {\bf 11}, 3056 (2020).

\bibitem{Smylie2024}
M. P. Smylie, Z. Islam, A. Glatz, G. D. Gu, J. A. Schneeloch, R. D. Zhong, S. Rosenkranz, W.-
K. Kwok, and U. Welp, Multimodal synchrotron x-ray diffraction across the superconducting
transition of Sr$_{0.10}$Bi$_2$Se$_3$ , Phys. Rev. B {\bf 110}, 024509 (2024).



\bibitem{WangZQ}
S. Zhou and Z. Wang,
Chern Fermipocket, topological pair density
wave, and charge-4e and charge-6e superconductivity in kagome superconductors,
Nat. Commun.       {\bf 13 },   (2022).



\bibitem{Fernandes-Fu}
R. M. Fernandes and L. Fu
Charge-4e Superconductivity from Multicomponent Nematic Pairing:
Application to Twisted Bilayer Graphene.
PHYSICAL REVIEW LETTERS {\bf 127}, 047001 (2021).

\bibitem{Grinenko}
I. Shipulin, N. Stegani, I. Maccari, K. Kihou,
C.-H. Lee, Q. Hu, Y. Zheng, F. Yang, Y. Li,
C.-M. Yim, R. Hühne, H.-H. Klauss,
F. Caglieris, E. Babaev,
M. Putti,
and  V. Grinenko,
Nat. Commun.       {\bf 14 }, 6734  (2023).

\bibitem{Kriener2011}
M. Kriener, K. Segawa, Z. Ren, S. Sasaki, S. Wada, S. Kuwabata, and Y. Ando, Electrochem-
ical synthesis and superconducting phase diagram of Cu$_x$Bi$_2$Se$_3$ , Phys. Rev. B {\bf 84}, 054513
(2011).



%\bibitem{Fu-PRB2014}
%L. Fu,  Odd-parity topological superconductor with nematic order: application to {\cx}3. Phys. Rev. B 90, 100509 (2014).

\bibitem{SM}
See Supplemental Materials for $c$-axis distortion and its disappearance  under a magnetic field above $H_{c2}$. %a summary of sample characterizations and the results for $x$=0.34.



\bibitem{Sun2019}
Y. Sun, S. Kittaka, T. Sakakibara, K. Machida, J. Wang, J. Wen, X. Xing, Z. Shi, and
T. Tamegai, Quasiparticle evidence for the nematic state above $T_c$ in Sr$_x$Bi$_2$Se$_3$, Phys. Rev.
Lett. {\bf 123}, 027002 (2019).

\bibitem{Hecker2018}
M. Hecker and J. Schmalian, Vestigial nematic order and superconductivity in the doped
topological insulator Cu$_x$Bi$_2$Se$_3$ , npj Quantum Materials {\bf 3}, 26 (2018).

\bibitem{Varlamov}
A. Larkin and A. Varlamov, {\it Theory of fluctuations in superconductors}, Oxford University Press, New York, (2005).

\bibitem{Arpes}
 E. Lahoud, E. Maniv, M. S. Petrushevsky, M. Naamneh, A. Ribak, S. Wiedmann, L. Petaccia, Z. Salman,
K. B. Chashka, Y. Dagan, and A. Kanigel, Evolution of the Fermi surface of a doped topological
insulator with carrier concentration. Phys. Rev. B {\bf 88}, 195107 (2013).

\end{thebibliography}

\end{document}